\documentclass[final,5p,times,twocolumn]{elsarticle}

\usepackage{amssymb}
\usepackage{amsmath}

\usepackage{graphicx}
\usepackage{bm}
\usepackage{slashed}
\usepackage{color}

\usepackage{amsfonts}
\usepackage{mathtools}
\usepackage{amstext}

\usepackage[colorlinks=true,pdfstartview=FitV,bookmarks=true,bookmarksnumbered=true,
breaklinks]{hyperref}
\hypersetup{linkcolor=red, citecolor=blue, urlcolor=cyan}

\usepackage[normalem]{ulem} 
\usepackage[dvipsnames]{xcolor} 

\journal{Physics Letters B}
\begin{document}

\begin{frontmatter}



\title{$J/\psi$-meson photoproduction off the nucleon in a dynamical model}

\author{Sang-Ho Kim} 
\affiliation{organization={Department of Physics and Origin of Matter and Evolution
of Galaxies (OMEG) Institute, Soongsil University}, 
            addressline={},
            city={Seoul},
            postcode={06978},
            state={},
            country={Republic of Korea}}

\begin{abstract}

The photoproduction of $J/\psi$ meson off the nucleon is investigated within a
dynamical model approach based on a Hamiltonian which describes the reaction
mechanisms of the Pomeron exchange, meson exchange, and direct $J/\psi$ radiation
terms.
To shed light on the low-energy mechanism, we scrutinize the role of light-meson
[$\pi^0(135)$, $\eta(548)$, $\eta'(958)$, $f_1(1285)$] and charmonium-meson
[$\eta_c(1S)$, $\chi_{c0}(1P)$, $\chi_{c1}(1P)$, $\eta_c(2S)$, $\chi_{c1}(3872)$]
exchanges in the $t$-channel diagram.
The values of the coupling constants are mostly determined by the radiative decays of
the $J/\psi$ meson or the relevant charmonium mesons.
The final $J/\psi$-$N$ interaction is required by the unitarity condition and is
described by the gluon-exchange and direct $J/\psi N$ coupling terms.
The parameters of the Hamiltonian are determined by the latest GlueX and $J/\psi$-007
experiments at the Jefferson Laboratory (JLab).
We find that $\eta(548)$ and $\eta'(958)$ light mesons give the most significant
contribution to $\gamma p \to J/\psi p$ among the meson-exchange terms.
Meanwhile, the effect of charmonium mesons turns out to be small compared to that of
light mesons.
It turns out that the contribution of the FSI term is about 1 - 2 orders of
magnitude smaller than that of the Born term and is significant near the threshold
when the Yukawa form is used as for the charmonium-nucleon potential.
The resulting total and $t$-dependent differential cross sections provide good
agreement with the JLab data.
The angle-dependent data with high precision near the very threshold ($E_\gamma
\leqslant 8.9$ GeV) are strongly desired to pin down the role of the FSI term more
properly.

\end{abstract}



\end{frontmatter}



\section{Introduction}
\label{sec1}

It is well known that the photoproduction of light vector mesons ($\rho$, $\omega$,
and $\phi$) and $J/\psi$ meson off the nucleon at high energies are governed by the
diffractive processes and are described by the $t$-channel Pomeron exchange.
In case of the production of light vector mesons, the slowly rising cross sections
with the beam energy $W$ is verified by the soft Pomeron exchange with its
trajectory $\alpha_{\mathbb P} (t) = 1.08 + 0.25t$~\cite{Laget:2000gj}.
Meanwhile, the measurements for $J/\psi$ meson reveal a steeper dependence on
$W$ than expected from the universal soft Pomeron exchange.
To account for the FermiLab~\cite{Binkley:1981kv,E687:1993hlm},
ZEUS~\cite{ZEUS:2002wfj}, and H1~\cite{H1:1996kyo,H1:2000kis} data in the region
$W \geqslant 10$ GeV, additional hard Pomeron exchange is suggested by Donnachie and
Landshoff (DL)~\cite{Donnachie:1998gm,Donnachie:1999qv,Donnachie:1999yb,
Donnachie:2001he}.

Other reaction models for $\gamma p \to J/\psi p$~\cite{Lee:2022ymp} involve the
two- and three-gluon exchanges based on perturbative quantum chromodynamics (pQCD)
and effective heavy quark field theory~\cite{Brodsky:2000zc}, the pQCD approach to
calculate the two-gluon exchange using the generalized parton distribution (GPD) of
the nucleon~\cite{Guo:2021ibg}, the exchanges of scalar ($0^{++}$) and tensor ($2^{++}$)
glueballs within the holographic formulation~\cite{Mamo:2019mka,Mamo:2021tzd}, etc.
The possibility of the existence of the $s$-channel hidden-charm pentaquark states
$P_c$ is also studied~\cite{Wang:2015jsa,Kubarovsky:2015aaa,Karliner:2015voa,
HillerBlin:2016odx,Wang:2019krd,Wu:2019adv,Winney:2019edt,JPAC:2023qgg}
since it was reported by the LHCb Collaboration in the $\Lambda_b^0 \to J/\psi K^- p$
decay~\cite{LHCb:2015yax,LHCb:2019kea}.

The author and his collaborators recently studied photo- and electroproduction of
$\phi$ meson within an effective Lagrangian approach~\cite{Kim:2019kef,Kim:2020wrd,
Kim:2021adl,Kim:2024lis} and found that some $t$-channel light mesons, such as
$\pi^0(135)$, $\eta(548)$, $a_0(980)$, $f_0(980)$, and $f_1(1285)$, play a crucial
role in describing the CEBAF Large Acceptance Spectrometer
(CLAS)~\cite{Seraydaryan:2013ija,Dey:2014tfa,Lukashin:2001sh,Santoro:2008ai} and
Laser Electron Photon Experiments at SPring-8 (LEPS)~\cite{LEPS:2010ovn} data at low
energies 2 $\leqslant$ $W$ $\leqslant$ 3 GeV although they are expected to be suppressed
due to the Okubo-Zwieg-Iizuka (OZI)
rule~\cite{Okubo:1963fa,Okubo:1963fa2,Okubo:1963fa3}.

A different situation occurs in case of the $J/\psi$-photoproduction mechanism.
The radiative decays of $J/\psi$ meson mostly proceed by producing multi-mesonic
resonant or non-resonant states.
Its radiative decays to the $\pi^0$ and $\eta$ mesons are about an order of magnitude
smaller than those of $\phi$ meson~\cite{PDG:2024cfk}.
Instead, as the mass scale incerases, some charmonium mesons, such as $\chi_{c0} (1P)$
and $\chi_{c1} (1P)$, have a large product of branching ratios to the $J/\psi \gamma$
channel~\cite{PDG:2024cfk} indicating that their exchanges could give non-negligible
contributions to the $t$-channel diagram.

In this letter, we extend our previous work for the $\phi$
photoproduction~\cite{Kim:2021adl,Kim:2024lis} to the production of $J/\psi$ meson
within the Hamiltonian formulation~\cite{Matsuyama:2006rp,Kamano:2019gtm}.
We construct a model Hamiltonian with the parameters determined by the recent
Jefferson Laboratory (JLab) data of $J/\psi$ photoproduction off the
nucleon~\cite{GlueX:2019mkq,GlueX:2023pev,Duran:2022xag}.
We follow the DL model~\cite{Donnachie:1984xq,Donnachie:1985iz,Donnachie:1987pu}
for the description of Pomeron exchange at high energies.
In the low-energy region, i.e., $4 \leqslant W \leqslant 5$ GeV, we elaborate on each
contribution of light mesons [$\pi^0(135)$, $\eta(548)$, $\eta'(958)$, $f_1(1285)$]
and charmonium mesons [$\eta_c(1S)$, $\chi_{c0} (1P)$, $\chi_{c1} (1P)$, $\eta_c (2S)$,
$\chi_{c1}(3872)$] in the $t$ channel.
The direct $J/\psi$ radiations from the nucleon are also considered in the $s$- and
$u$-channels.

The unitarity condition requires the $\gamma N \to J/\psi N$ amplitude to include
the $J/\psi N \to J/\psi N$ final-state interaction (FSI) as well.
The $J/\psi N \to J/\psi N$ reaction is essential for exploring the possible
$J/\psi$-nucleus bound states predicted by lattice QCD (LQCD)
calculations~\cite{Beane:2014sda}.
As for the possible sources of the $J/\psi N$ interaction, we consider the
gluon-exchange mechanism within QCD and the direct $J/\psi N$
coupling amplitudes.

In the next section, we explain the general formalism of our dynamical model of
$J/\psi$ photoproduction off the nucleon.
In sections~\ref{sec3} and~ \ref{sec4}, the formalism of the Born and FSI terms is
explained, respectively, in detail.
In section~\ref{sec5}, we present the numerical results for the total and
$t$-dependent differential cross sections and discuss them in comparison with the
higher-statistics JLab data.
Section.~\ref{sec6} is devoted to summary and conclusion.


\section{Dynamical model of the $\gamma p \to J/\psi p$ reaction}
\label{sec2}

We can define a model Hamiltonian which can generate the $\gamma N \to J/\psi N$
reaction and $J/\psi N \to J/\psi N$ FSI term following the dynamical
formulation~\cite{Matsuyama:2006rp,Kamano:2019gtm}
\begin{align}
H =& H_0 +B_{J/\psi N,\gamma N} + \Gamma_{N^*,\gamma N} + \Gamma_{N^*, J/\psi N},
\label{eq:model-h}
\end{align}
where $H_0$ is the free Hamiltonian of the system and $B_{J/\psi N,\gamma N}$ is the Born
amplitude consisting of the tree diagrams for $\gamma N \leftrightarrow J/\psi N$.
The last two terms indicate the nucleon resonance ($N^*$) contribution in the $s$
channel~\cite{Wang:2015jsa,Kubarovsky:2015aaa,Karliner:2015voa,HillerBlin:2016odx,
Wang:2019krd,Wu:2019adv,Winney:2019edt,JPAC:2023qgg}
and are not considered in this letter.

\begin{figure}[h]
\begin{center}
\includegraphics[width=8.8cm]{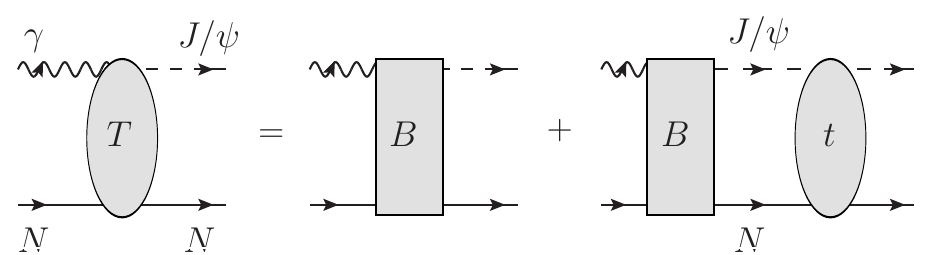}
\caption{Full amplitude for the $\gamma N \to J/\psi N$ reaction:
$B$ is the Born amplitude and $t$ is the $J/\psi N \to J/\psi N$ scattering amplitude.}
\label{FIG01}
\end{center}
\end{figure}

\begin{figure}[h]
\centering
\includegraphics[width=8.8cm]{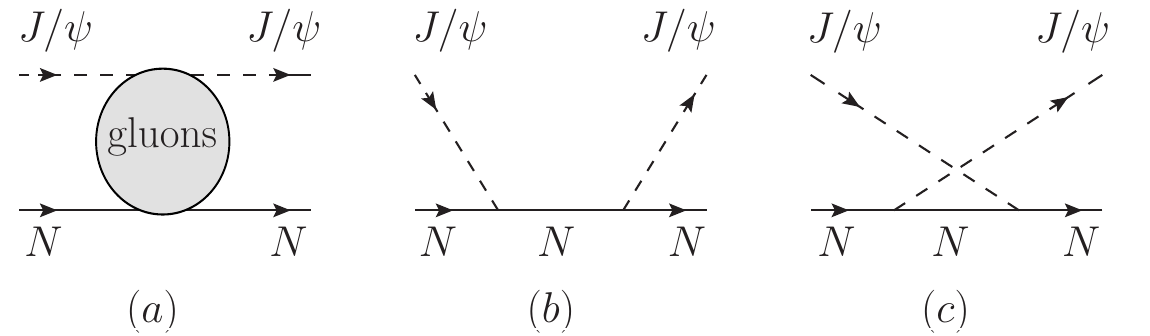}
\caption{$J/\psi N \to J/\psi N$ scattering amplitudes:
(a) gluon-exchange within QCD, (b,c) direct $J/\psi N$ coupling terms.}
\label{FIG02}
\end{figure}

The full amplitude for $\gamma N \to J/\psi N$ defined by the Hamiltonian
of Eq.~(\ref{eq:model-h}) is illustrated in Fig.~\ref{FIG01} and can be written as
\begin{align}
T_{J/\psi N,\gamma N}(E) =  B_{J/\psi N,\gamma N} + T^{\rm FSI}_{J/\psi N,\gamma N}(E), 
\label{eq:t-gn-phin}
\end{align}
where the last term denotes the $J/\psi N$ FSI amplitude and can be expressed as
\begin{align}
T^{\rm FSI}_{J/\psi N,\gamma N}(E) = t_{J/\psi N, J/\psi N}(E)\, G_{J/\psi N}(E)\,
B_{J/\psi N,\gamma N}.
\label{eq:t-fsi}
\end{align}
Here the meson-baryon propagator is given by
\begin{align}
G_{MB}(E)=\frac{|\, MB\, \rangle\, \langle\, MB\, |}{E-H_0+i\epsilon},
\label{eq:mb-prop}
\end{align}
and the $J/\psi N \to J/\psi N$ scattering amplitude is defined by
\begin{align}
& t_{J/\psi N, J/\psi N}(E) 
\cr &
= V_{J/\psi N, J/\psi N}(E) +
V_{J/\psi N, J/\psi N}(E) \, G_{J/\psi N}(E) \, t_{J/\psi N, J/\psi N}(E) ,
\label{eq:lseq}
\end{align}
where the $J/\psi N$ potential is decomposed into
\begin{align}
V_{J/\psi N, J/\psi N}(E) =
v_{J/\psi N, J/\psi N}^{\rm Gluon} (E) + v_{J/\psi N, J/\psi N}^{\rm Direct} (E),
\label{eq:VNpot}
\end{align}
as illustrated in Fig.~\ref{FIG02}.
Here $v_{J/\psi N, J/\psi N}^{\rm Gluon}$ indicates the gluon-exchange interaction
[Fig.~\ref{FIG02}(a)] and $v_{J/\psi N, J/\psi N}^{\rm Direct}$ the direct $J/\psi N$
coupling term [Figs.~\ref{FIG02}(b,c)].

With the normalization condition
$\langle \mbox{\bf k} \vert \mbox{\bf k}' \rangle =
\delta^3 (\mbox{\bf k}-\mbox{\bf k}')$
for plane wave states~\cite{GW} and
$\langle \psi_B \vert \psi_B \rangle = 1$ for a single particle state $\psi_B$,  
the differential cross section of
\begin{align}
\gamma(q,\lambda_\gamma) + N(p_i,m_s) \to J/\psi(k,\lambda_{J/\psi}) + N(p_f,m'_s),
\end{align}
in the center of mass (c.m.) frame, can be written as
\begin{align}
\frac{d\sigma}{d\Omega} =& \frac{(2\pi)^4}{q^2} \rho_{J/\psi N} (W) \rho_{\gamma N} (W)
\frac14 \sum_{\lambda_{J/\psi},m_s'} \sum_{\lambda_\gamma,m_s}
\cr & \times
\left| \langle
k \lambda_{J/\psi} ; p_fm_s' \vert T_{J/\psi N, \gamma N} (W) \vert q \lambda_\gamma ; p_i m_s
\rangle \right|^2,
\end{align}
where
\begin{align}
\rho_{J/\psi N} = \frac{kE_{J/\psi} (\mbox{\bf k})E_N(\mbox{\bf k})}{W},\,\,\,
\rho_{\gamma N} = \frac{q^2 E_N (\mbox{\bf q})}{W},
\end{align}
with $\mbox{\bf p}_i = -\mbox{\bf q}$ and $\mbox{\bf p}_f = -\mbox{\bf k}$.
The invariant mass is given by
$W = q + E_N (\mbox{\bf q}) = E_{J/\psi}(\mbox{\bf k}) + E_N(\mbox{\bf k})$.
$\lambda_\gamma$ and $\lambda_{J/\psi}$ are the helicities of the photon and $J/\psi$
meson, respectively,
and $m_s$ and $m_s'$ are the magnetic quantum numbers of the initial and final
nucleons, respectively.

\begin{figure*}[ht]
\centering
\includegraphics[width=16.0cm]{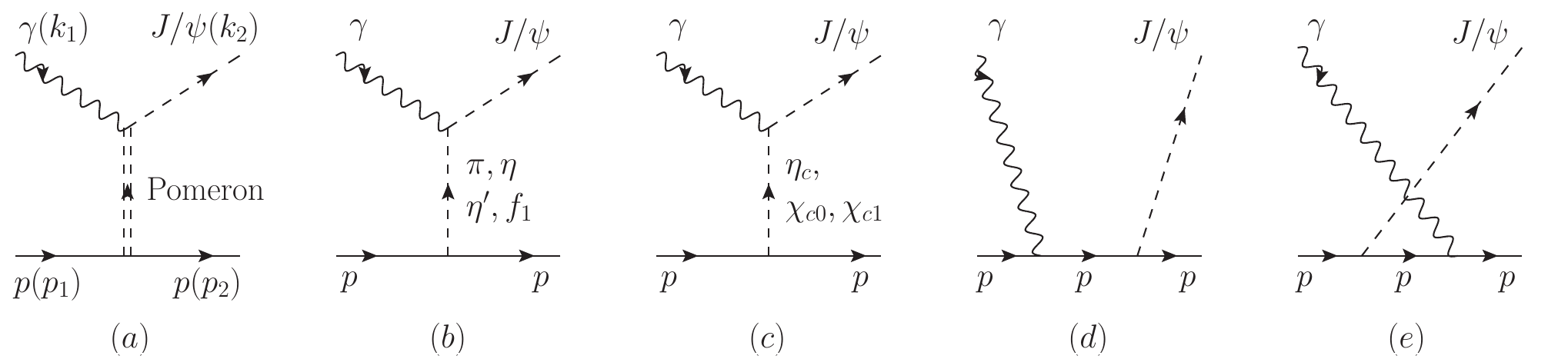}
\caption{Born diagrams for $\gamma (q) + p (p_i) \to J/\psi (k) + p (p_f)$,
which include the exchanges of ($a$) Pomeron, ($b$) light mesons [$\pi^0$, $\eta$,
$\eta'$, $f_1(1285)$], ($c$) charmonium mesons [$\eta_c(1S)$, $\chi_{c0}(1P)$,
$\chi_{c1}(1P)$, $\eta_c (2s)$, $\chi_{c1}(3872)$] in the $t$ channel, $(d,\,e)$
direct $J/\psi$ radiation terms in the $s$- and $u$-channels.}
\label{FIG03}
\end{figure*}

\section{Born term}
\label{sec3}

The Feynman diagrams of the Born term for $\gamma p \to J/\psi p$ are drawn in
Fig.~\ref{FIG03}, which include the exchanges of the ($a$) Pomeron, ($b$) light
mesons, ($c$) charmonium mesons in the $t$ channel, and ($d,\,e$) direct $J/\psi$
radiations in the $s$- and $u$-channels.
The scattering amplitude for the Born term reads
\begin{align}
& \langle k \lambda_{J/\psi}; p_fm_s' \mid B_{J/\psi N, \gamma N} \mid q\lambda_\gamma ;
p_i m_s \rangle
\cr
= & \frac{1}{(2\pi)^3} \sqrt{ \frac{M_N^2}{4 E_{J/\psi}(\mbox{\bf k})
E_N(\mbox{\bf p}_f) \left\vert \mbox{\bf q} \right\vert E_N(\mbox{\bf p}_i)} }
\cr & \times
\left[ \bar{u}_N (p_f,m'_s) \mathcal{M}^{\mu\nu}(k,p_f,q,p_i) u_N (p_i,m_s) \right]
\cr & \times
\epsilon^*_\nu(k,\lambda_{J/\psi}) \epsilon_\mu (q,\lambda_\gamma) ,
\end{align}
where $\epsilon_\mu(q,\lambda_\gamma)$ is the photon polarization vector with momentum
$q$ and helicity $\lambda_\gamma$, and $\epsilon_\nu(k,\lambda_{J/\psi})$ is that of the
$J/\psi$ meson with momentum $k$ and helicity $\lambda_{J/\psi}$.
The nucleon spinor of momentum $p$ and spin projection $m_s$ is given by
$u_N (p,m_s)$, which is normalized as $\bar{u}_N (p,m_s) u_N (p,m'_s) =
\delta_{m_s,m'_s}$.
$E_{J/\psi}$ and $E_N$ are the energies of the $J/\psi$ meson and nucleon,
respectively.
Each term of Fig.~\ref{FIG03} will be explained in detail below.

\subsection{Pomeron exchange}

In the DL model~\cite{Donnachie:1984xq,Donnachie:1985iz,Donnachie:1987pu}, the
two-gluon exchange mechanism is parametrized as the soft Pomeron exchange within the
Regge phenomenology [Fig.~\ref{FIG03}(a)].
The scattering amplitude takes the form
\begin{align}
\mathcal{M}_{\mathbb{P}}^{\mu\nu} = i\frac{12eM_V^2}{f_V}
\beta_{q_V} F_{V}(t) \beta_{u/d} F_N(t)
[ \rlap{\,/}{q}g^{\mu\nu} - \gamma^\mu q^\nu ]
G_{\mathbb{P}}(s,t),
\label{eq:Amp:SPom}
\end{align}
where $M_h$ is the mass of a hadron $h$.
$\beta_{q_V}$ ($\beta_{u/d}$) indicates the coupling between the Pomeron and the
quark in a vector meson $V$ (nucleon $N$) and is determined by fitting to the
cross section data of the diffractive photoproduction of vector mesons:
\begin{align}
\beta_{u/d} = 2.07,\,\,\, \beta_s = 1.39,\,\,\, \beta_c = 0.323\,[{\rm GeV}^{-1}].
\end{align}

To get the vector meson decay constant $f_V$, we recall that the vector meson
dominance (VMD) is defined by the effective Lagrangian of
\begin{align}
{\mathcal L}_{\mathrm{VMD}} = \frac{e M_V^2}{f_V} A^\mu V_\mu,
\label{eq:VMD-0}
\end{align}
where $A^\mu$ and $V_\mu$ are the fields of the photon and vector meson,
respectively.
Then, $f_V$ can be calculated from the experimental data of the width of
$V \to e^+e^-$~\cite{PDG:2024cfk} and
\begin{align}
\Gamma_{V \to e^+ e^-} = \alpha^2 \frac{4\pi}{3} \frac{M_V}{f_V^2} ,
\label{eq:VMD:DW}
\end{align}
where $\alpha = e^2/(4\pi)$ with $e$ being the unit electric charge.
We finally obtain
\begin{align}
& f_\rho = 4.94,\,\,\, f_\omega = 17.06,   \cr
& f_\phi = 13.38,\,\,\, f_{J/\psi} = 11.18,\,\,\, f_{\Upsilon} = 39.68.
\end{align}

The propagator $G_{\mathbb{P}}$ in Eq.~(\ref{eq:Amp:SPom}) is an important
ingredient of the Regge phenomenology and takes the form
\begin{align}
G_{\mathbb{P}} (s,t) =
\left( \frac{s}{s_{\mathbb P}} \right)^{\alpha_{\mathbb P}(t)-1}
\mathrm{exp} \left\{ -\frac{i\pi}{2} [\alpha_{\mathbb P}(t)-1] \right\}
\label{eq:Prop:SPom}.
\end{align}
The Pomeron trajectory is determined to be $\alpha_{\mathbb{P}}(t) = 1.25 + 0.25t$
and the mass scale is given by
$s_{\mathbb{P}} = 1/\alpha_{\mathbb P}' = 4\,\mathrm{GeV}^2$.
The form factor for the Pomeron-vector meson vertex and the isoscalar
electromagnetic (EM) form factor of the nucleon are, respectively, defined by
\begin{align}
F_V (t) =& \frac{2 \mu_0^2}{(M_V^2-t)(2\mu_0^2+M_V^2-t)}, \\
F_N(t) =& \frac{4M^2_N-2.8t}{(4M^2_N-t)(1-t/0.71)^2} ,
\label{eq:FF:PomExch}
\end{align}
where $\mu_0 = 1.1\,{\rm GeV}^2$ and $t = (q - k)^2$.

\subsection{Light-meson exchanges}

\begin{table*}[h]
\centering
\begin{tabular}{cccccc}
\hline
Mesons\,($M$)&Mass ($J^P$)\hspace{1.5em}&$\Gamma_M$ [MeV]
&\hspace{3.0em}$\mathrm{Br}_{J/\psi \to M\gamma}\,[\%]$\hspace{2.0em}
&$|g_{\gamma M J/\psi}|$
&\hspace{2em}$g_{MNN}$\hspace{2em} \\
\hline
$\pi$&134 ($0^-$)&$-$&(3.39 $\pm$ 0.08) $\cdot$ $10^{-3}$&
1.83 $\cdot$ $10^{-3}$&13.0 \\
$\eta$&548 ($0^-$)&1.31 $\cdot$ $10^{-3}$ &(1.090 $\pm$ 0.013) $\cdot$ $10^{-1}$&
1.09 $\cdot$ $10^{-2}$ &6.34 \\
$\eta'$&958 ($0^-$)&0.188&(5.28 $\pm$ 0.06) $\cdot$ $10^{-1}$&
2.65 $\cdot$ $10^{-2}$ &6.87 \\
$f_1$&1285 ($1^+$)&23.0&(6.1 $\pm$ 0.8) $\cdot$ $10^{-2}$&
3.93 $\cdot$ $10^{-3}$&2.5 \\
$\eta_c(1S)$&2984 ($0^-$)&30.5 &1.41 $\pm$ 0.14&
1.95 & - \\
\hline
\end{tabular}
\caption{Radiative decays of $J/\psi$ into the light mesons and
$\eta_c(1S)$~\cite{PDG:2024cfk} and strong coupling constants
$g_{MNN}$~\cite{Stoks:1999bz,Stoks:1999bz2}.}
\label{TAB1}
\end{table*}

\begin{table*}[h]
\centering
\begin{tabular}{ccccccc}
\hline
Mesons ($M$)&\hspace{1.5em}Mass ($J^P$)\hspace{1.5em}&$\Gamma_M$ [MeV]
&\hspace{1.0em}$\mathrm{Br}_{M\to J/\psi\gamma}\,[\%]$\hspace{1.0em}
&$|g_{\gamma M J/\psi}|$
&\hspace{3.0em}$\mathrm{Br}_{M\to  p \bar p}\,[\%]$\hspace{2.0em}
&$|g_{Mpp}|$ \\
\hline
$\eta_c(1S)$ &2984 ($0^-$)& 30.5 & - & -
&(1.33 $\pm$ 0.11) $\cdot$ $10^{-1}$& 2.70 $\cdot$ $10^{-2}$ \\
$\chi_{c0} (1P)$&3414 ($0^+$)&10.7&1.41 $\pm$ 0.09&1.33
&(2.21 $\pm$ 0.14) $\cdot$ $10^{-2}$& 4.56 $\cdot$ $10^{-3}$ \\
$\chi_{c1} (1P)$&3511 ($1^+$)&0.84&34.3 $\pm$ 1.3&3.29
&(7.6 $\pm$ 0.4) $\cdot$ $10^{-3}$&8.42 $\cdot$ $10^{-4}$ \\
$\eta_c (2S)$&3638 ($0^-$)&11.8&$<$ 1.4&$<$ 1.32
&$<$ 2.0 $\cdot$ $10^{-1}$ &$<$ 1.61 $\cdot$ $10^{-2}$ \\
$\chi_{c1}(3872)$&3872 ($1^+$)&1.19
&0.78 $\pm$ 0.29 & 0.257
&$<$ 2.2 $\cdot$ $10^{-3}$&$<$ 5.11 $\cdot$ $10^{-4}$ \\
\hline
\end{tabular}
\caption{Radiative decays of the charmonium mesons into $J/\psi$ and strong coupling
constants $g_{Mpp}$~\cite{PDG:2024cfk}.}
\label{TAB2}
\end{table*}

First, in case of light-meson exchanges [Fig.~\ref{FIG03}(b)], we consider
($\pi,\,\eta,\,\eta'$) pseudoscalar mesons and $f_1(1285)$ axial-vector meson into
which the radiative decays of $J/\psi$ are non-negligible as shown in
Table.~\ref{TAB1}.

The effective Lagrangians for the EM interaction read
\begin{align}
\mathcal{L}_{\gamma \Phi J/\psi} =& \frac{e g_{\gamma \Phi J/\psi}}{M_{J/\psi}}
\epsilon^{\mu\nu\alpha\beta} \partial_\mu A_\nu \partial_\alpha \psi_\beta \Phi,
\label{eq:Lag:GMJP-1} \\
\mathcal{L}_{\gamma A_1 J/\psi} =& \frac{e g_{\gamma A_1 J/\psi}}{M_{J/\psi}^2}
\epsilon^{\mu\nu\alpha\beta} \partial_\mu A_\nu
\partial^\delta \partial_\delta \psi_\alpha A_{1\beta},
\label{eq:Lag:GMJP-2}
\end{align}
where $\Phi$, $\psi_\beta$, and $A_{1\beta} $ are the fields of the pseudoscalar,
$J/\psi$, and axial-vector mesons, respectively.
The coupling constants are calculated from the widths of the $J/\psi \to \Phi
\gamma$ and $J/\psi \to A_1 \gamma$ radiative decays
\begin{align}
\Gamma_{J/\psi \to \Phi \gamma} =&
\frac{e^2}{12\pi} \frac{q_\gamma^3}{M_{J/\psi}^2} g_{\gamma \Phi J/\psi}^2,
\label{eq:Wid:GMJP-1} \\
\Gamma_{J/\psi \to A_1 \gamma} =&
\frac{e^2}{12\pi} \frac{M_{J/\psi}^2+M_{A_1}^2}{M_{J/\psi}^2 M_{A_1}^2} q_\gamma^3
g_{\gamma A_1 J/\psi}^2,
\label{eq:Wid:GMJP-2}
\end{align}
where $q_\gamma = (M_{J/\psi}^2- M_{\Phi,\,A_1}^2)/(2M_{J/\psi})$.
The decay width of $J/\psi$ is given by $\Gamma_{J/\psi} = 92.6$ KeV.

The strong interaction Lagrangians are written as
\begin{align}
\mathcal{L}_{\Phi NN} =& -ig_{\Phi NN} \bar N \Phi \gamma_5 N ,
\label{eq:Lag:MNN-1} \\
\mathcal{L}_{A_1 NN} =& - g_{A_1 NN} \bar N
\left[ \gamma_\mu - i \frac{\kappa_{A_1 NN}}{2M_N} \gamma_\nu \gamma_\mu
\partial^\nu \right] A_1^\mu \gamma_5 N,
\label{eq:Lag:MNN-2}
\end{align}
where $g_{\Phi NN}$ is obtained using the Nijmegen
potentials~\cite{Stoks:1999bz,Stoks:1999bz2}.
$g_{f_1 NN}$ is taken from Refs.~\cite{Birkel:1995ct,Kochelev:1999zf} and
the tensor term $\kappa_{f_1 N N}$ is set to be zero for brevity.
We summarize all the relevant parameters in Table~\ref{TAB1}.

\subsection{Charmonium-meson exchanges}

Second, in case of charmonium-meson exchanges, [Fig.~\ref{FIG03}(c)], we consider
$\eta_c(1S)$, $\chi_{c0}(1P)$, and $\chi_{c1}(1P)$ mesons.
$\eta_c(2S)$ and $\chi_{c1}(3872)$ mesons also give sizable upper limits for their
branching ratios to $J/\psi \gamma$ or $p \bar p$.
Thus we test their contributions as well.

For the exchange of $\chi_{c0}(1P)$ scalar meson, the EM interaction can be expressed
by the following effective Lagrangian
\begin{align}
\mathcal{L}_{\gamma S J/\psi} =& \frac{eg_{\gamma S J/\psi}}{M_{J/\psi}}
F^{\mu\nu} \psi_{\mu\nu} S,
\label{eq:Lag:GSJP}
\end{align}
where $S$ is the field of the scalar meson.
$F^{\mu\nu}$ and $\psi_{\mu\nu}$ are the EM and $J/\psi$ field-strength tensors,
respectively:
$F^{\mu\nu} = \partial^\mu A^\nu - \partial^\nu A^\mu$ and
$\psi_{\mu\nu} = \partial_\mu \psi_\nu - \partial_\nu \psi_\mu$.
$g_{\gamma S J/\psi}$ can be obtained from the radiative decay width of
$S \to J/\psi \gamma$
\begin{align}
\Gamma_{S \to J/\psi \gamma} =
\frac{e^2}{\pi} \frac{q_\gamma^3}{M_{J/\psi}^2} g_{\gamma S J/\psi}^2,
\label{eq:Wid:GSJP}
\end{align}
where $q_\gamma = (M_S^2- M_{J/\psi} ^2)/(2M_S)$.

When $\eta_c(1S)$ pseudoscalar exchanges is considered, we use the same formulae of
Eq.~(\ref{eq:Lag:GMJP-1}) and (\ref{eq:Wid:GMJP-1}) because of
$M_{J/\psi} > M_{\eta_c (1S)}$.
For the exchanges of $\eta_c(2S)$ pseudoscalar and [$\chi_{c1}(1P)$,\,$\chi_{c1}(3872)$]
axial-vector mesons, the effective Lagrangians are the same as for those of
Eqs.~(\ref{eq:Lag:GMJP-1}) and (\ref{eq:Lag:GMJP-2}), respectively.
The coupling constants are also obtained from the radiative decays of
$\Phi \to J/\psi \gamma$ and $A_1 \to J/\psi \gamma$,
the formula of which are given by
\begin{align}
\Gamma_{\Phi \to J/\psi \gamma} =&
\frac{e^2}{3\pi} \frac{q_\gamma^3}{M_{J/\psi}^2} g_{\gamma \Phi J/\psi}^2,
\label{eq:Wid:GMJP-3} \\
\Gamma_{A_1 \to J/\psi \gamma} =&
\frac{e^2}{12\pi} \frac{M_{J/\psi}^2+M_{A_1}^2}{M_{J/\psi}^2 M_{A_1}^2} q_\gamma^3
g_{\gamma A_1 J/\psi}^2,
\label{eq:Wid:GMJP-4}
\end{align}
where $q_\gamma = (M_{\Phi,\,A_1}^2-M_{J/\psi}^2)/(2M_{\Phi,\,A_1})$.

The effective Lagrangian of the scalar meson with the nucleon takes the form
\begin{align}
\mathcal{L}_{S NN} = -g_{S NN} \bar N S N.
\label{eq:Lag:SNN}
\end{align}
The decays of the relevant charmonium mesons into $p \bar p$ are known
experimentally~\cite{PDG:2024cfk}.
Thus the strong coupling constants for the exchanges of the pseudoscalar,
axial-vector, and scalar charmonium mesons are obtained from their decays into
$p \bar p$
\begin{align}
\Gamma_{\Phi \to p \bar p} =&
\frac{1}{\pi} \frac{q_N^3}{M_\Phi^2} g_{\Phi p p}^2,
\label{eq:Wid:MNN-1} \\
\Gamma_{A_1 \to p \bar p} =&
\frac{1}{3\pi} \frac{q_N}{M_{A_1}^2} (3M_N^2+2q_N^2)
g_{A_1 p p}^2,
\label{eq:Wid:MNN-2} \\
\Gamma_{S \to p \bar p} =&
\frac{1}{\pi} \frac{q_N}{M_S^2}(M_N^2+q_N^2) g_{S pp}^2,
\label{eq:Wid:MNN-3}
\end{align}
which are derived from Eqs.~(\ref{eq:Lag:MNN-1}), (\ref{eq:Lag:MNN-2}), and
(\ref{eq:Lag:SNN}), respectively.
Here $q_N = (M_{\Phi,\,A_1,\,S}^2- 4M_N^2)^{\frac12}/2$.
All the relevant parameters are tabulated in Table.~\ref{TAB2}.
The parameters relevant to $J/\psi \to \eta_c(1S) \gamma$ are given in
Table.~\ref{TAB1} for easy comparison.

The scattering amplitudes for the pseudoscalar, axial-vector, and scalar meson
exchanges read
\begin{align}
\mathcal{M}_\Phi^{\mu\nu} =& \frac{ie}{M_{J/\psi}}
\frac{g_{\gamma \Phi J/\psi} g_{\Phi NN}}{t-M_\Phi^2} \epsilon^{\mu\nu\alpha\beta} q_{\alpha}
k_{\beta} \gamma_5,
\label{eq:MesonEachAmpl-1} \\
\mathcal{M}_{A_1}^{\mu\nu} =& i e
\frac{g_{\gamma A_1 J/\psi} g_{A_1 NN}}{t - M_{A_1}^2}
\epsilon^{\mu\nu\alpha\beta}
\left( -g_{\alpha\lambda}+\frac{q_{t\alpha} q_{t\lambda}}{M_{A_1}^2} \right)
\cr &\times
\left( \gamma^\lambda + \frac{\kappa_{A_1 NN}}{2M_N} \gamma^\sigma \gamma^\lambda
q_{t\sigma} \right) \gamma_5 q_\beta,
\label{eq:MesonEachAmpl-2} \\
\mathcal{M}_S^{\mu\nu} =& - \frac{2e}{M_{J/\psi}}
\frac{g_{\gamma S J/\psi} g_{S NN}}{t- M_S^2}
(q \cdot k g^{\mu\nu} - q^\mu k^\nu),
\label{eq:MesonEachAmpl-3}
\end{align}
respectively, where $q_t = k - q$.
For the exchanged mesons with finite width $\Gamma_M$, the mass is replaced by
$M_M \to (M_M -i \Gamma_M/2)$ in the propagator.

To preserve the unitarity condition, we use the Regge prescription for
$\mathcal{M}_{f_1} = \mathcal{M}_{A_1}$ where
the Feynman propagator is replaced with the Regge one as
\begin{align}
\frac{1}{t - M_{f_1}^2} \to
\left(\frac{s}{s_{f_1}} \right)^{\alpha_{f_1}(t)-1}
\frac{\pi\alpha'_{f_1}}{\sin[\pi\alpha_{f_1}(t)]}
\frac{1}{\Gamma [\alpha_{f_1}(t)]} D_{f_1}(t).
\label{eq:f1:Prop}
\end{align}
Here $s_{f_1} = 1$~GeV$^2$ and $\alpha_{f_1} = 0.95 + 0.028 t$~\cite{Kochelev:1999zf}.
The signature factor is in the form of~\cite{Kochelev:1999zf}
\begin{align}
D_{f_1}(t) = \frac{{\rm exp}[-i\pi\alpha_{f_1}(t)]-1}{2}.
\label{eq:f1:SigFac}
\end{align}

Each scattering amplitude of Eqs.~(\ref{eq:MesonEachAmpl-1})-(\ref{eq:MesonEachAmpl-3})
is multiplied by the form factor given by
\begin{align}
F_M (t) = \frac{\Lambda_M^4}{\Lambda_M^4 + (t - M_M^2)^2},
\label{eq:FF:M}
\end{align}
where the cutoff parameters are determined as
$(\Lambda_\Phi,\,\Lambda_S,\,\Lambda_{A_1}) = 2.1$~GeV.




\subsection{Direct $J/\psi$ radiations}

It is found that the direct $\phi$ radiations play a crucial role in $\phi$
photoproduction~\cite{Kim:2019kef,Kim:2021adl,Kim:2024lis}.
Thus we include the direct $J/\psi$ radiation term also in $J/\psi$ photoproduction,
as drawn in Figs.~\ref{FIG03}(d,e).
The relevant effective Lagrangians are defined by
\begin{align}
\mathcal{L}_{\gamma NN} =& - e \bar N
\left[ \gamma_\mu - \frac{\kappa_N}{2M_N} \sigma_{\mu\nu} \partial^\nu
\right] N A^\mu ,
\label{eq:Lag:GammaNN} \\
\mathcal{L}_{J/\psi NN} =& - g_{J/\psi NN} \bar N
\left[ \gamma_\mu - \frac{\kappa_{J/\psi NN}}{2M_N} \sigma_{\mu\nu} \partial^\nu
\right] N \psi^\mu,
\label{eq:Lag:JPNN}
\end{align}
where the anomalous magnetic moment of the proton is given by
$\kappa_p = 1.79$.
We set the tensor term of the $J/\psi NN$ vertex to be zero for simplicity.
The coupling constant in the vector term is obtained from the branching ratio for
$J/\psi \to p \bar p$ and
\begin{align}
\Gamma_{J/\psi \to p \bar p} =
\frac{2}{3\pi} \frac{q_N^3}{M_{J/\psi}^2} g_{J/\psi p p}^2,
\label{eq:Wid:JPNN}
\end{align}
where $q_N = (M_{J/\psi}^2- 4M_N^2)^{\frac12}/2$.
From $\Gamma_{J/\psi} = 92.6$ KeV and
$\rm Br_{J/\psi \to p \bar p} = 2.120 \cdot 10^{-3}$~\cite{PDG:2024cfk}, we obtain
$|g_{J/\psi N N}| = |g_{J/\psi p p}| = 2.18 \cdot 10^{-3}$.

The $J/\psi$ radiation scattering amplitudes for $s$- and $u$-channels read
\begin{align}
\mathcal{M}_{J/\psi,{\rm rad},s}^{\mu\nu} &= \frac{e g^{}_{J/\psi NN}}{s-M_N^2}
\left( \gamma^\nu - i\frac{\kappa_{J/\psi NN}}{2M_N} \sigma^{\nu\alpha} k_{\alpha}
\right)
\nonumber \\ & \quad \times 
(\slashed{q}_s + M_N)
\left( \gamma^\mu + i\frac{\kappa_N}{2M_N} \sigma^{\mu\beta} q_{s \beta} \right),
\label{eq:Ampl:N-1} \\
\mathcal{M}_{J/\psi,{\rm rad},u}^{\mu\nu} &= \frac{e g^{}_{J/\psi NN}}{u-M_N^2}
\left( \gamma^\mu + i\frac{\kappa_N}{2M_N} \sigma^{\mu\alpha} q_{u \alpha} \right)
\nonumber \\ & \quad \times
(\slashed{q}_u + M_N)
\left( \gamma^\nu - i\frac{\kappa_{J/\psi NN}}{2M_N} \sigma^{\nu\beta} k_{\beta} \right),
\label{eq:Ampl:N-2}
\end{align}
respectively, where $q_s = q + p_i$ and $q_u = p_f - q$.

We choose the following form factor
\begin{align}
F_N (\mbox{\bf k}) = \frac{\Lambda_N^2}{\Lambda_N^2 + \mbox{\bf k}^2},
\label{eq:FF:JPNN}
\end{align}
to make sure that the integration in calculating the FSI term converges.
$\mbox{\bf k}$ is the 3-momentum of the produced $J/\psi$ meson.
The cutoff parameter is chosen to be $\Lambda_N = 2.1$ GeV.
Note that similar amplitudes are needed for considering the FSI effect through
$J/\psi p \to J/\psi p$ as shown in Figs.~\ref{FIG02}(b,c) where the same form
factor of Eq.~(\ref{eq:FF:JPNN}) will be used.

\section{FSI term} \label{sec:fsi}
\label{sec4}

The amplitude for the FSI is defined in Eqs.~(\ref{eq:t-fsi})-(\ref{eq:VNpot}). 
We keep only the leading term in Eq.~(\ref{eq:lseq}) which leads to the
approximation $t_{J/\psi N, J/\psi N}(E) \sim V_{J/\psi N, J/\psi N}(E)$ in evaluating the
FSI amplitude in Eq.~(\ref{eq:t-fsi}) such that the individual contributions of
$v^{\rm Gluon}_{J/\psi N, J/\psi N}$ and $v^{\rm Direct}_{J/\psi N, J/\psi N}$ can be studied.
The amplitude  $T^{\rm FSI}_{J/\psi N, \gamma N}(E)$ of Eq.~(\ref{eq:t-fsi}) for the
reaction of
$\gamma (\mbox{\bf q}) + N(-\mbox{\bf q}) \to
J/\psi(\mbox{\bf k}) + N(-\mbox{\bf k})$
can then be written as 
\begin{align}
& \langle \mbox{\bf k} \vert T^{\rm FSI}_{J/\psi N, \gamma N}(E) \vert \mbox{\bf q}
\rangle
=\int d \mbox{\bf k}'
\langle \mbox{\bf k} \vert V_{J/\psi N, J/\psi N}(E) \vert \mbox{\bf k}' \rangle
\cr & \mbox{} \times
\frac{1}{E-E_{J/\psi} (k')-E_N(k')+i\epsilon}
\langle \mbox{\bf k}' \vert B_{J/\psi N, \gamma N} \vert \mbox{\bf q} \rangle,
\label{eq:fsi-born}
\end{align}
in the c.m. frame.

Now, we need to evaluate the matrix elements of the potential of
$V_{J/\psi N,J/\psi N}(E)$ for $J/\psi (k') + N (p') \to J/\psi (k) + N(p)$, which
can be expressed as
\begin{align}
& \langle k \lambda_{J/\psi}; p m_s \vert V_{J/\psi N, J/\psi N} \vert k'\lambda'_{J/\psi};
p' m'_s \rangle
\cr  = &
\frac{1}{(2\pi)^3}\sqrt{\frac{ M_N^2 }{4 E_{J/\psi} (\mbox{\bf k}) E_N(\mbox{\bf p})
E_{J/\psi}(\mbox{\bf k}') E_N(\mbox{\bf p}^{'})}}
\cr & \times
\mathcal{V}(k\lambda_{J/\psi}, p m_s; k' \lambda'_{J/\psi}, p' m'_s),
\end{align}
where
\begin{align}
\mathcal{V}=\mathcal{V}_{\rm Gluon} + \mathcal{V}_{\rm Direct}.
\end{align}
Here $\mathcal{V}_{\rm Gluon}$ [Fig.~\ref{FIG02}(a)] and $\mathcal{V}_{\rm Direct}$
[Figs.~\ref{FIG02}(b,c)] represent the gluon-exchange interaction and direct
$J/\psi N$ coupling terms, respectively.
More details will be followed for these $J/\psi N$ potentials from the interaction
Lagrangians by using the unitary transformation method of the ANL-Osaka
formulation~\cite{Matsuyama:2006rp,Kamano:2019gtm}.

\subsection{Gluon exchange interaction}

Because of the OZI rule, the $J/\psi N$ interaction is expected to be governed by
gluon exchanges.
Since there exists the LQCD calculations for the $J/\psi N$ potential, we use the
form suggested by Ref.~\cite{Kawanai:2010ev} for the charmonium-nucleon potential
which is found to be approximately of the Yukawa form.
We therefore take the form of
\begin{align}
\mathcal{V}_{\rm Gluon}=-v_0 \frac{e^{-\alpha r}}{r},
\label{eq:yukawa}
\end{align}
which was also adopted in the phenomenological $c \bar c$-$N$
potential~\cite{Brodsky:1989jd}.
The parameters in Eq.~(\ref{eq:yukawa}) are given
by~\cite{Kawanai:2010ev,Brodsky:1989jd}
\begin{align}
v_0 = 0.1,\,\,\, \alpha = 0.6\, {\rm GeV}, \\
v_0 = 0.4,\,\,\, \alpha = 0.6\, {\rm GeV}.
\label{eq:yukawa-2}
\end{align}
We attempt to use both of them in our calculation and compare the results each
other in the next section.

The potential of Eq.~(\ref{eq:yukawa}) can be obtained by taking the nonrelativistic
limit of the scalar meson exchange amplitude calculated from the Lagrangian
\begin{align}
\mathcal{L}_{\sigma} = V_0 \left( \bar{\psi}_N\psi_N \Phi_\sigma +
\psi^\mu \psi_\mu \Phi_\sigma \right) ,
\label{eq:lag-s}
\end{align}
where $\Phi_\sigma$ is a scalar field with mass $\alpha$ in Eq.~(\ref{eq:yukawa}). 
By using the unitary transformation method~\cite{Matsuyama:2006rp,Kamano:2019gtm},
the scalar-meson exchange matrix element derived from Eq.~(\ref{eq:lag-s}) can be
written as
\begin{align}
&\mathcal{V}_{\rm Gluon}(k\lambda_{J/\psi},p m_s;k'\lambda'_{J/\psi},p'm'_s) = 
\frac{V_0}{(p-p')^2 - \alpha^2}
\cr &\,\times
[\bar{u}_N(p,m_s)\, u_N(p',m'_s)]
[\epsilon^*_\mu(k,\lambda_{J/\psi})\, \epsilon^\mu (k',\lambda'_{J/\psi})],
\label{eq:yukawa-rel}
\end{align}
where $V_0 = -8 v_0 \pi M_{J/\psi}$
and ${(p - p')} = ( E_N(p)-E_N(p'),\,\mbox{\bf p}-\mbox{\bf p}')$.

\subsection{Direct $J/\psi N$ coupling term}

The form of the direct $J/\psi N$ coupling amplitudes is the same as given in
Eqs.~(\ref{eq:Ampl:N-1}) and (\ref{eq:Ampl:N-2}) after the replacement of $e$ by
$g_{J/\psi NN}$ and $\kappa_N$ by $\kappa_{J/\psi  N N}$.
We also use the same form of the form factor as given in Eq.~(\ref{eq:FF:JPNN}) and
the same cutoff parameter, i.e., $\Lambda_N = 2.1$~GeV.

\begin{figure}[t]
\centering
\includegraphics[width=7.5cm]{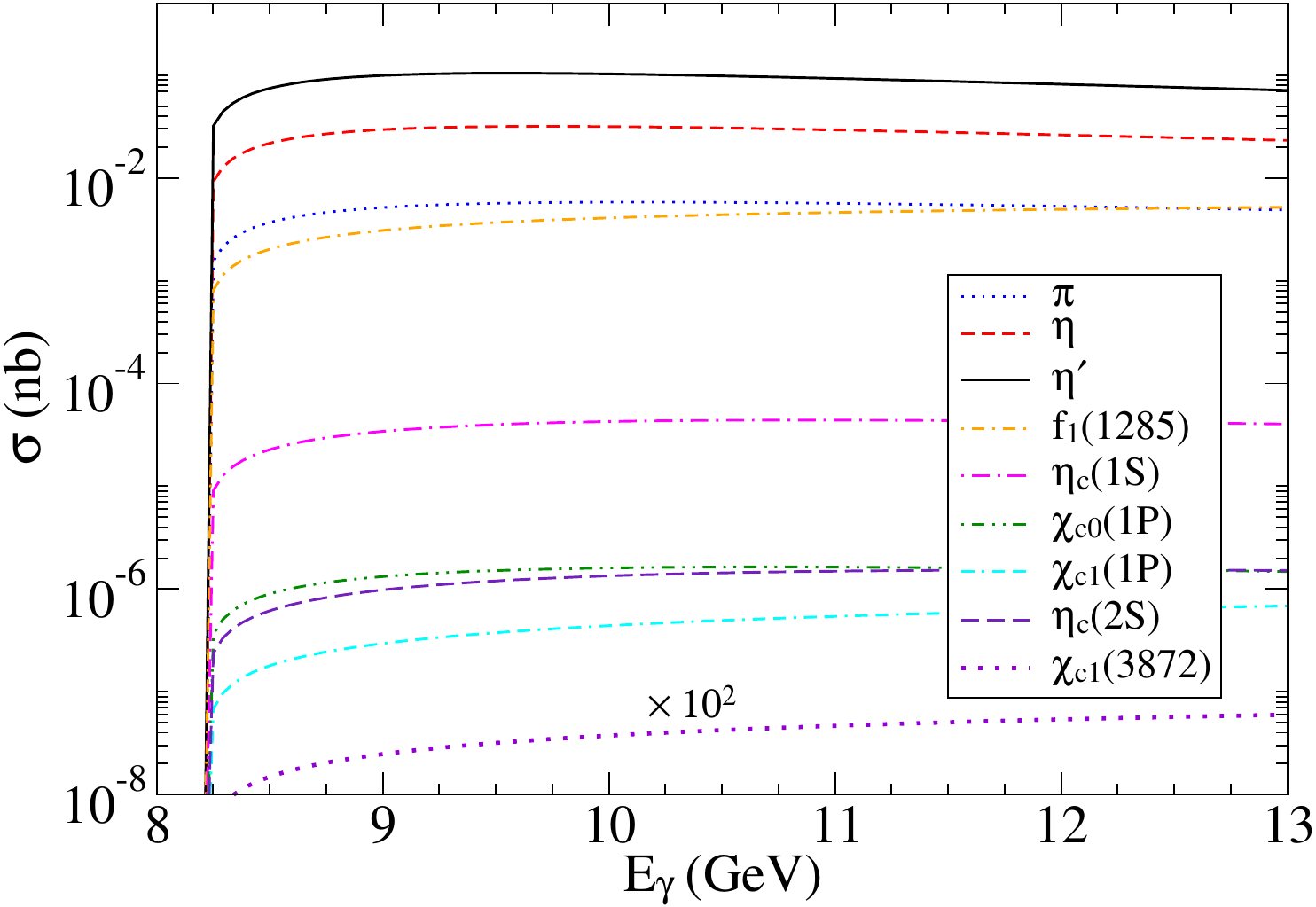}
\caption{Each contribution of meson exchanges for $\gamma p \to J/\psi p$ is plotted
as a function of the Lab energy $E_\gamma$.}
\label{FIG04}
\end{figure}

\section{Numerical Results and Discussions}
\label{sec5}

\subsection{Born term}

\begin{figure}[h!] 
\centering
\includegraphics[width=7.5cm]{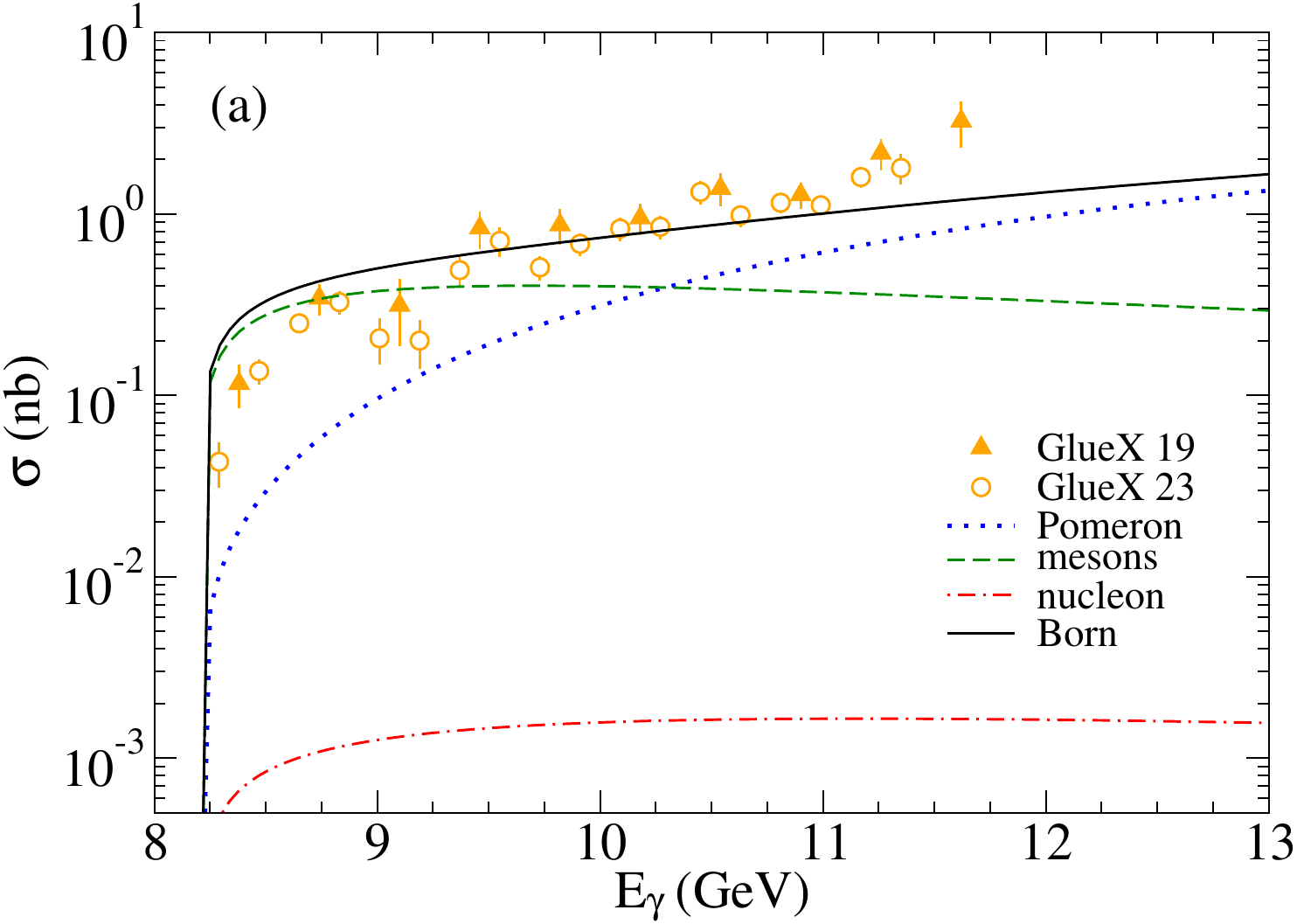} \\ \vspace{1em}
\includegraphics[width=7.5cm]{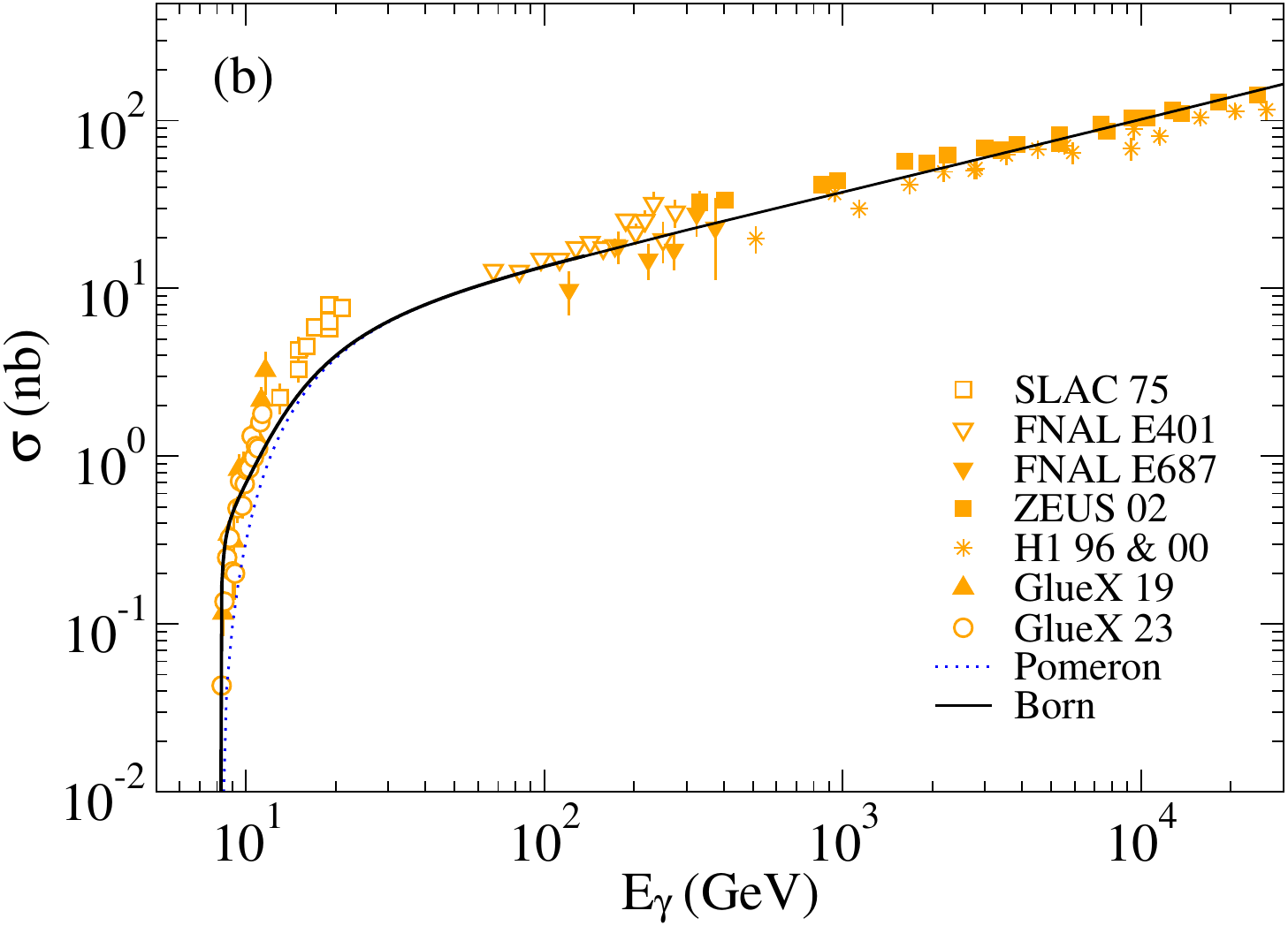}
\caption{(a) Total cross section for $\gamma p \to J/\psi p$ without FSI is plotted
as a function of the Lab energy $E_\gamma$ at low energies ($8 \leqslant E_\gamma
\leqslant 13$ GeV). 
The blue dotted, green dashed, and red dot-dashed curves stand for the contributions
of Pomeron, meson, and nucleon exchanges, respectively.
The black solid curve denote the Born contribution.
(b) Total cross section from threshold up to $E_\gamma \simeq 10^4$ GeV.
The SLAC~\cite{Camerini:1975cy}, FermiLab~\cite{Binkley:1981kv,E687:1993hlm},
ZEUS~\cite{ZEUS:2002wfj}, H1~\cite{H1:1996kyo,H1:2000kis}, and
GlueX~\cite{GlueX:2019mkq,GlueX:2023pev} data are used.}
\label{FIG05}
\end{figure}

\begin{figure}[h!] 
\vspace{1em}
\centering
\includegraphics[width=7.5cm]{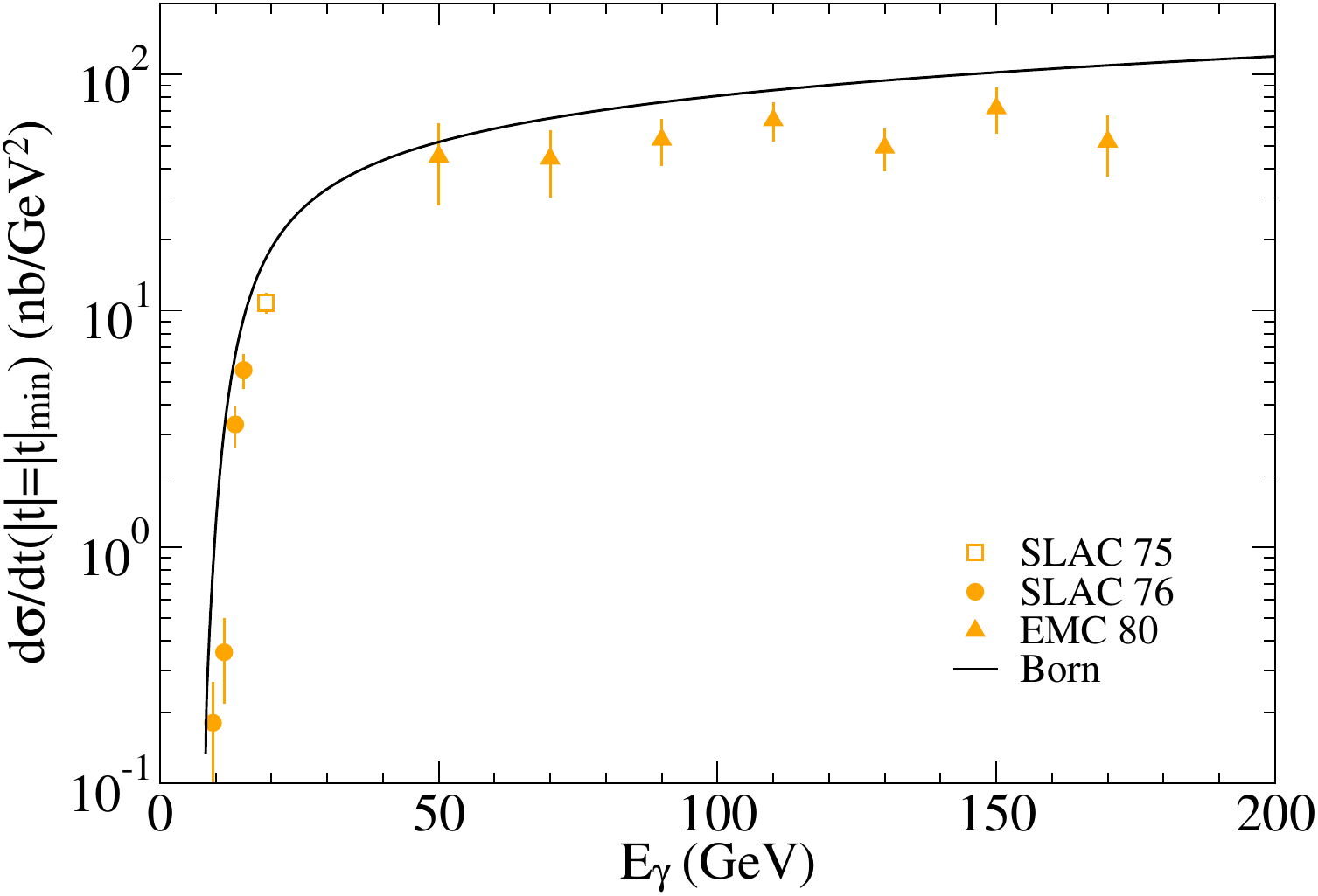}
\caption{Differential cross section is plotted as a function of the Lab energy
$E_\gamma$ at $|t| = |t|_{\rm min}$ from threshold up to $E_\gamma = 200$ GeV.
The SLAC~\cite{Camerini:1975cy,Ritson:1976rj} and EMC~\cite{EuropeanMuon:1979nky}
data are compared with the result from the Born contribution.}
\label{FIG06}
\end{figure}

\begin{figure*}[h]
\centering
\includegraphics[width=15cm]{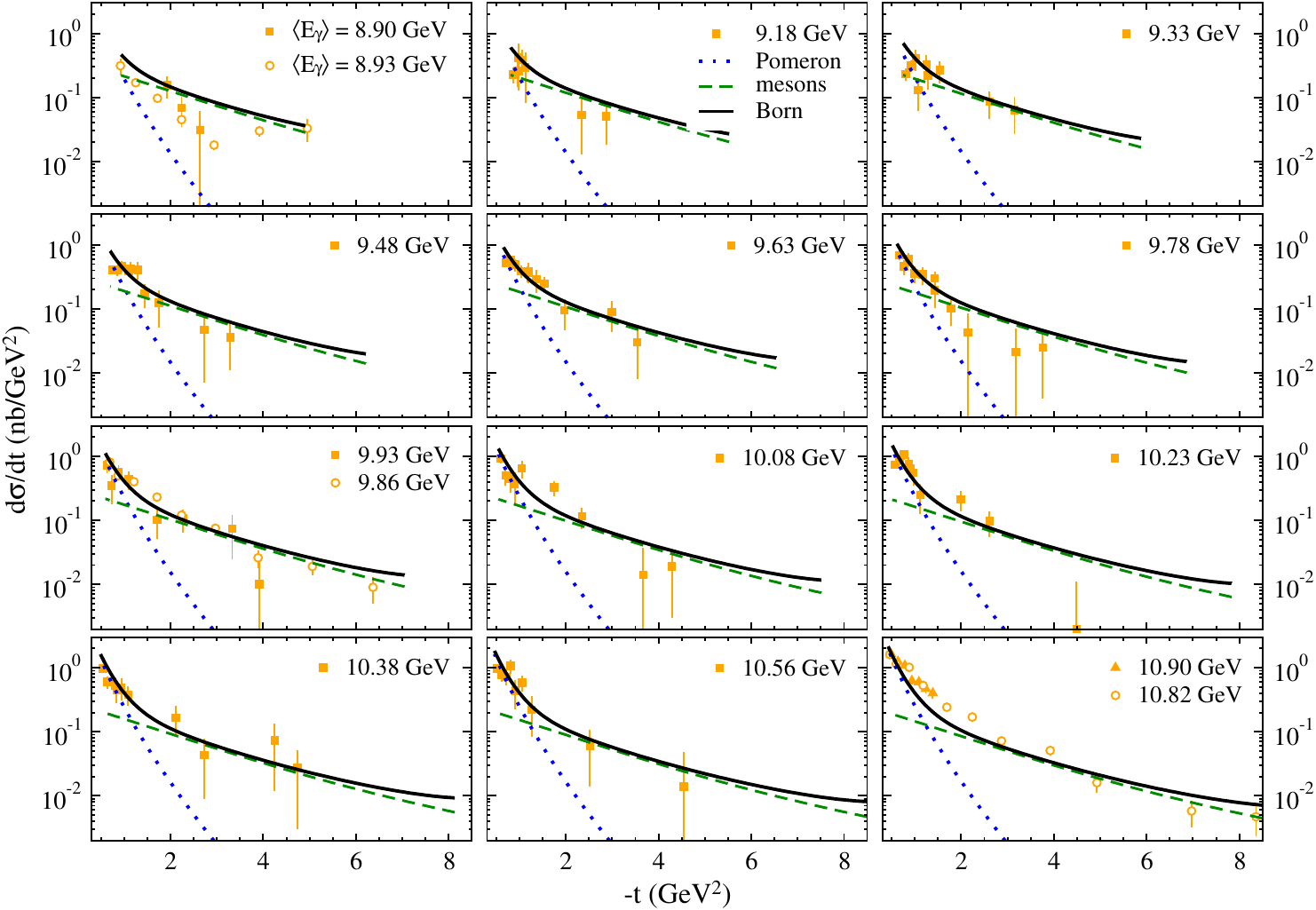}
\caption{Differential cross sections without FSI are plotted as functions of the
momentum transfer $-t$ at different Lab energies ($8.90 \leqslant E_\gamma
\leqslant 10.90$ GeV).
The GlueX19~\cite{GlueX:2019mkq} (triangle), GlueX23~\cite{GlueX:2023pev} (circle),
and $J/\psi$-007~\cite{Duran:2022xag} (quadrangle) data from the JLab are compared
with the Pomeron-exchange (blue dotted), meson-exchange (green dashed), and Born
(black solid) contributions.}
\label{FIG07}
\end{figure*}

We first show the results when the Born term is only considered as discussed in
Sec.~\ref{sec3}.
To examine the $t$-channel meson-exchange mechanism, each contribution of light-meson
and charmonium-meson exchanges is displayed in Fig.~\ref{FIG04} as a function of
the photon Lab energy $E_\gamma$.
In case of $\eta_c(2S)$ and $\chi_{c1}(3872)$ mesons, only the upper limits for their
branching ratios to $J/\psi \gamma$ or $p \bar p$ are provided.
Thus their contributions also indicate upper limits on the total cross section.
We find that the differences of the cross sections between the light-meson and
charmonium-meson contributions are large enough to be distinguished.
Clearly, light mesons make more significant contribution than charmonium mesons.
The most dominant contribution from the light- and charmonium-mesons turns out to be
$\eta'(958)$ and $\eta_c(1S)$ mesons, respectively.
Nevertheless, $\eta_c(1S)$ exchange is about three orders of magnitude suppressed
relative to $\eta'(958)$ one.

Figure~\ref{FIG05}(a) depicts the total cross section of the Born contribution
(black solid) which involves the Pomeron exchange (blue dotted) and the sum of meson
exchanges (green dashed) shown in Fig.~\ref{FIG04}.
The direct $J/\psi$-radiation contribution (red dot-dashed) is also given.
The results indicate that the contribution of meson exchanges is indeed essential for
describing the low-energy GlueX data~\cite{GlueX:2019mkq,GlueX:2023pev} together with
that of Pomeron exchange.
Meanwhile, the direct $J/\psi$-radiation contribution is more suppressed than the
meson-exchange one by a factor of about $10^2$.
In Fig.~\ref{FIG05}(b), we show the contributions of the Pomeron exchange and Born
term from threshold up to $E_\gamma \simeq 10^4$ GeV.
The Pomeron exchange describes the high energy data~\cite{Binkley:1981kv,
E687:1993hlm,ZEUS:2002wfj,H1:1996kyo,H1:2000kis,Camerini:1975cy} quite well.

\begin{figure*}[h] 
\centering
\includegraphics[width=7.5cm]{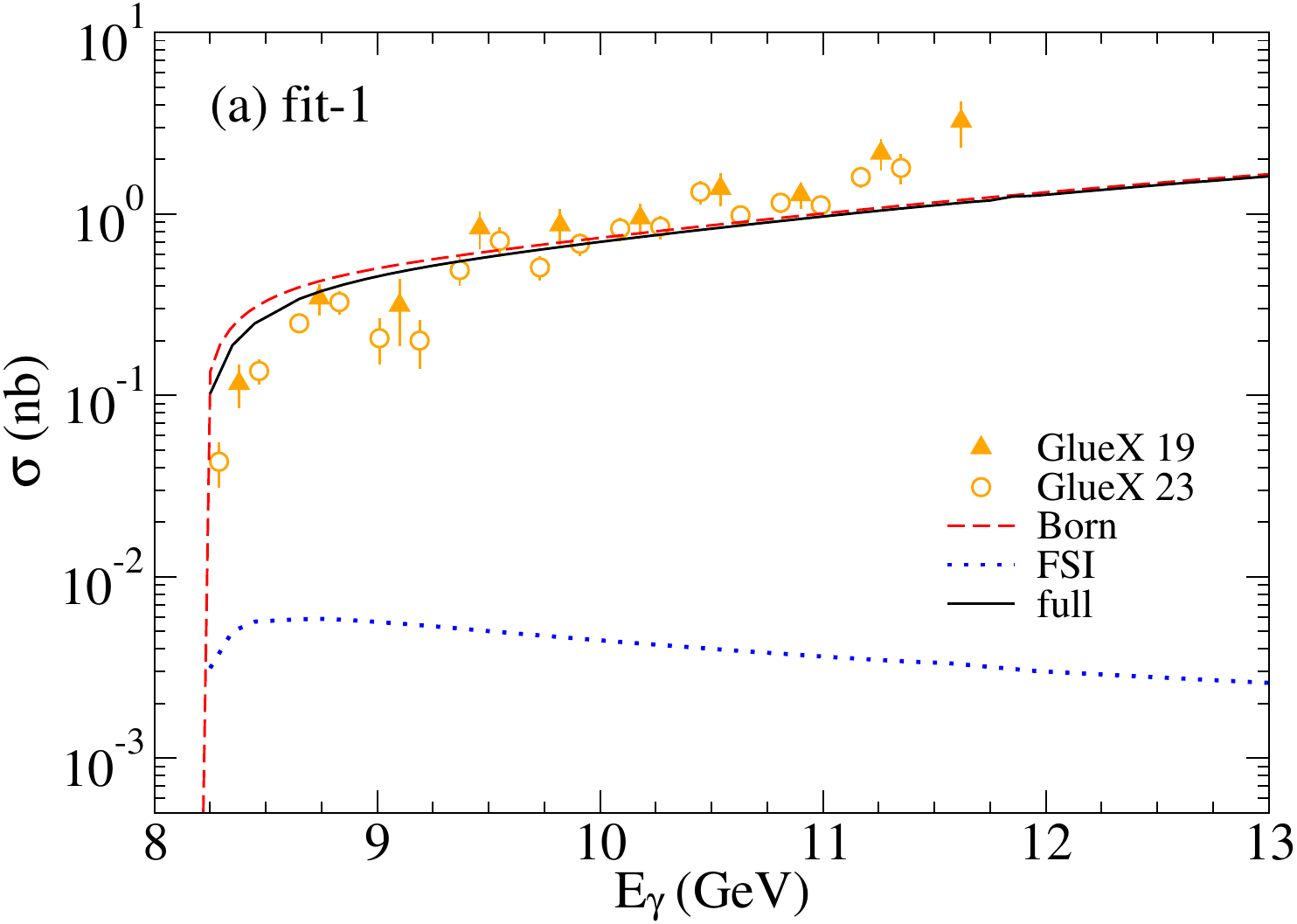} \hspace{1em}
\includegraphics[width=7.5cm]{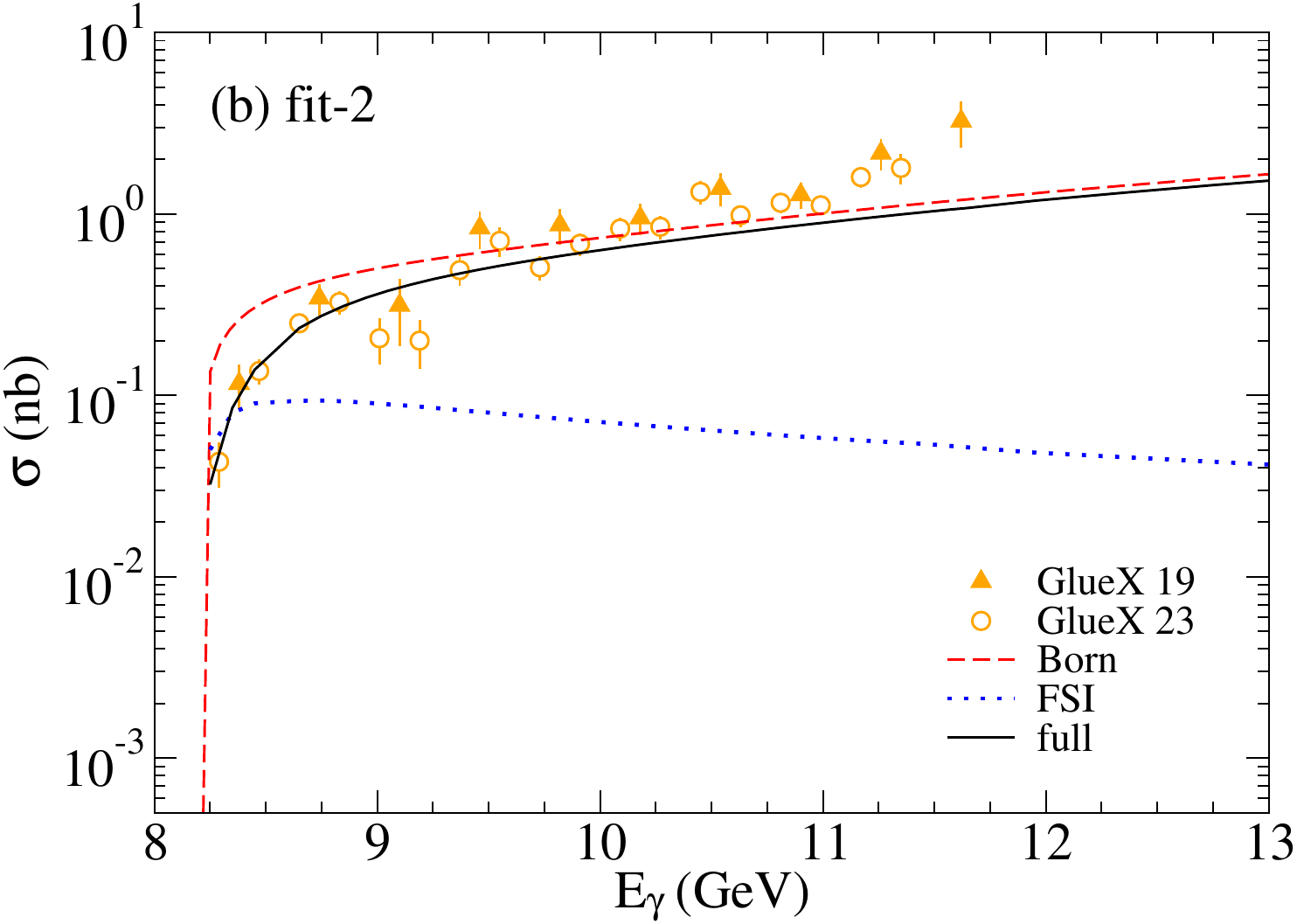}
\caption{Total cross section for $\gamma p \to J/\psi p$ with FSI is plotted as a
function of the Lab energy $E_\gamma$ at low energies ($8 \leqslant E_\gamma \leqslant
13$ GeV). 
The red dashed and blue dotted curves stand for the contributions of the Born and
FSI terms, respectively.
The black solid curve denote the full contribution.
The parameters of the Yukawa potential ($v_0$, $\alpha$) we have used:
(a) fit-1: (0.1, 0.6)~\cite{Kawanai:2010ev},
(b) fit-2: (0.4, 0.6)~\cite{Brodsky:1989jd}.
The data are from the GlueX Collaboration~\cite{GlueX:2019mkq,GlueX:2023pev}.}
\label{FIG08}
\end{figure*}

\begin{figure*}[h!]
\vspace{1em}
\centering
\includegraphics[width=15cm]{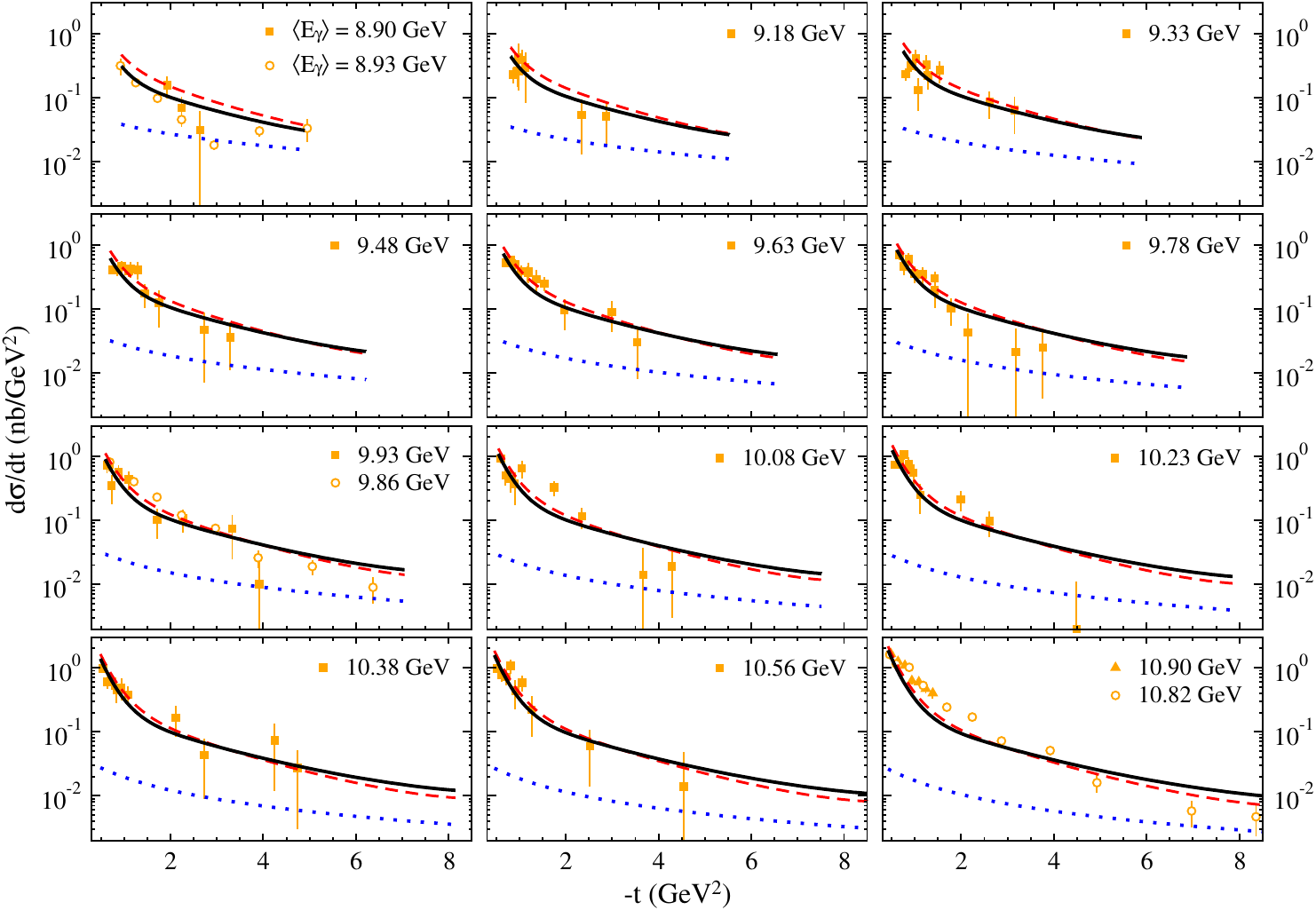}
\caption{Differential cross sections are plotted as functions of the momentum transfer
$-t$ at different Lab energies ($8.90 \leqslant E_\gamma \leqslant 10.90$ GeV)
with and without the FSI effect.
The GlueX19~\cite{GlueX:2019mkq} (triangle), GlueX23~\cite{GlueX:2023pev} (circle),
and $J/\psi$-007~\cite{Duran:2022xag} (quadrangle) data from the JLab are compared
with the results with the fit-2 model.
The curve notations are the same as those in Fig.~\ref{FIG08}.}
\label{FIG09}
\end{figure*}

In Fig.~\ref{FIG06}, we present the differential cross section as a function of the
Lab energy $E_\gamma$ at $|t| = |t|_{\rm min}$.
The SLAC~\cite{Camerini:1975cy,Ritson:1976rj} and European Muon Collaboration
(EMC)~\cite{EuropeanMuon:1979nky} data are compared with the result of the Born term.
The overall shape of the result is very similar to the available data although the
result slightly overestimates the experimental data.
Thus our Pomeron model is verified again by this comparison.

Figure~\ref{FIG07} displays the results for the differential cross sections as
functions of $-t$.
Pomeron exchange alone is insufficient to account for the low energy JLab
data~\cite{GlueX:2019mkq,GlueX:2023pev,Duran:2022xag}.
The inclusion of meson exchanges greatly improves the results over the whole
scattering angles because the slope of the meson-exchange contribution is very similar
to that of the JLab data at $-t \geqslant 2$ ${\rm GeV}^2$.
A similar conclusion is found in $\phi$-meson
photoproduction~\cite{Kim:2019kef,Kim:2021adl,Kim:2024lis} where light mesons, such as
$\pi^0(135)$, $\eta(548)$, $a_0(980)$, $f_0(980)$, and $f_1(1285)$, played an important
role for describing the available angle-dependent
data~\cite{Seraydaryan:2013ija,Dey:2014tfa}.
The direct $J/\psi$-radiation term begins to come into play in the backward
scattering regions as expected.
That is, because of its contribution, the Born term is increased to some extent in the
regions $-t \geqslant 5$ ${\rm GeV}^2$ in comparison with the meson-exchange term.
The GlueX23 data at $E_\gamma =$ (8.93, 9.86, 10.82) GeV~\cite{GlueX:2023pev} cover the
whole scattering angles and thus place constraint on the direct $J/\psi$-radiation
contribution.

\subsection{FSI term}

We are now in a position to show the results with the FSI term where the
gluon-exchange interaction and direct $J/\psi$ coupling terms are involved as
discussed in Sec.~\ref{sec4}.
It is found that the contribution of the FSI term entirely comes from the
gluon-exchange interaction.
Meanwhile the direct $J/\psi$ coupling term is relatively far more suppressed.
We attempt to use two different sets of parameters ($v_0$, $\alpha$) as for the
Yukawa potential of Eq.~(\ref{eq:yukawa}) in the gluon-exchange interaction:
(0.1, 0.6)~\cite{Kawanai:2010ev} and (0.4, 0.6)~\cite{Brodsky:1989jd}, which we call
as fit-1 and fit-2 models.

The results for the total cross section with the FSI effect are depicted in
Fig.~\ref{FIG08} where the red dashed curve (Born term) corresponds to the black
solid curve in Fig.~\ref{FIG05}.
It turns out that the the Born term interferes destructively with the FSI term.
The FSI effect is small when we use the fit-1 model (Fig.~\ref{FIG08}(a)) but is
significant when the fit-2 model is used (Fig.~\ref{FIG08}(b)).
Indeed, we find a noticeable improvement with the fit-2 model near the threshold.

The differential cross sections corresponding to the fit-2 model are displayed in
Fig.~\ref{FIG09} as functions of $-t$.
The contribution of the FSI term is about $10^1$ - $10^2$ times smaller than that of
the Born term depending on the kinematical regions.
The Born term interferes destructively with the FSI term at forward scattering angles
($-t \leqslant 4$ GeV$^2$) and interferes constructively at backward scattering angles
($-t \geqslant 4$ GeV$^2$).
The FSI term makes the result much better at $E_\gamma = 8.93$ GeV in particular.
More accurate angle-dependent data near the very threshold ($E_\gamma \leqslant 8.9$
GeV) are strongly desired to verify the role of the FSI term.
Note that both the direct $J/\psi$-radiation term in Born amplitude and the FSI term
enhance the results in the backward scattering regions ($-t \geqslant 4$ GeV$^2$).

\section{Summary and Conclusion}
\label{sec6}
In this letter, we aimed at investigating $J/\psi$-meson photoproduction off the
nucleon target within a dynamical model approach based on a
Hamiltonian~\cite{Matsuyama:2006rp,Kamano:2019gtm} which can generate the $\gamma N
\to J/\psi N$ reaction and $J/\psi N \to J/\psi N$ FSI term.
The Born amplitude for $\gamma p \to J/\psi p$ consists of the Pomeron exchange,
meson exchanges in the $t$ channel, and direct $J/\psi$ radiations in the $s$- and
$u$-channels.

The role of light-meson and charmonium-meson exchanges is extensively studied and the
relevant electromagnetic coupling constants involved in the effective Lagrangians are
determined by the radiative decays of $J/\psi$ and charmonium
mesons~\cite{PDG:2024cfk}.
Meanwhile, the meson-$NN$ coupling constants are taken from the Nijmegen potentials
and radiative decays of charmonium mesons to $p \bar p$ for the light- and
charmonium-meson exchanges, respectively.
We have found that $\eta(548)$ and $\eta'(958)$ light mesons play a crucial role to
describe the available JLab data~\cite{GlueX:2019mkq,GlueX:2023pev,Duran:2022xag} but
charmonium mesons give a negligible contribution to the $\gamma p \to J/\psi p$
reaction.

The final $J/\psi$-$N$ interaction is required by the unitarity condition and the
Yukawa form of [$-v_0\,\rm{exp}(-\alpha r)/r$] is assumed as for the
charmonium-nucleon potential when considering the gluon-exchange interaction.
To check the model dependence of the FSI effect, two different sets of parameters are
attempted~\cite{Kawanai:2010ev,Brodsky:1989jd}.
The FSI effect turned out to be essential to account for the threshold region when
the values of ($v_0$,\,$\alpha$) = (0.4,\,0.6)~\cite{Brodsky:1989jd} are used.
Note that the importance of the FSI effect is also found in
Ref.~\cite{Sakinah:2024cza} where a model based on the constituent quark model (CQM)
and a phenomenological charm quark-nucleon potential is constructed to investigate the
$\gamma p \to J/\psi p$ reaction.
Meanwhile, the author and his collaborators have found that the FSI effect is small in
$\phi$-meson photoproduction where a similar dynamical approach is
employed~\cite{Kim:2021adl}.

Our reaction model provides an excellent description of the total and $t$-dependent
differential cross sections and can be justified after more high-precision
angle-dependent data are accumulated near the very threshold ($E_\gamma \leqslant 8.9$
GeV).
Also the measurements of various spin polarization obervables by future experiments
will become valuable information for a better understanding of the $J/\psi$
photoproduction mechanism.


\section*{Acknowledgments}

The author is grateful to T.-S.~H.~Lee and S.-i.~Nam for valuable discussions.
The work was supported by the Basic Science Research Program through the National
Research Foundation of Korea (NRF) under Grants
No. RS-2021-NR060129 and
No. RS-2022-NR074739.




\end{document}